\documentclass[a4paper,fleqn]{cas-dc}
\usepackage[authoryear,longnamesfirst]{natbib}

\def\tsc#1{\csdef{#1}{\textsc{\lowercase{#1}}\xspace}}
\tsc{WGM}
\tsc{QE}

\newcommand{\Nc}{N_{\mathrm{c}}}
\newcommand{\Nf}{N_{\mathrm{f}}}
\newcommand{\GeV}{\,\mathrm{GeV}}
\newcommand{\mq}{m_{\mathrm{q}}}
\newcommand{\Tc}{T_{\mathrm{c}}}
\newcommand{\LQCD}{\Lambda_{\mathrm{QCD}}}
\newcommand{\muB}{\mu_{\mathrm{B}}}

\begin{document}
\let\WriteBookmarks\relax
\def\floatpagepagefraction{1}
\def\textpagefraction{.001}

\shorttitle{QCD Phase Diagram and Astrophysical Implications}    
\shortauthors{K.~Fukushima}
\title [mode = title]{QCD Phase Diagram and Astrophysical Implications}  
\author{Kenji Fukushima}[orcid=0000-0003-0899-740X]
\ead{fuku@nt.phys.s.u-tokyo.ac.jp}
\affiliation{organization={Department of Physics, The University of Tokyo},
            addressline={7-3-1 Hongo, Bunkyo-ku}, 
            city={Tokyo},
            postcode={113-0033}, 
            country={Japan}}

\begin{abstract}
I make a brief review about the QCD phases and the equation of state
inferred from the neutron star data.  Along the temperature axis at
low baryon density, the QCD phase transition is a smooth crossover,
and it is a natural extension of our imagination to postulate a
similar crossover along the density axis at low temperature.  Even
without phase transitions, the inferred thermodynamic properties of
neutron star matter turn out to be highly nontrivial already at twice
of the nuclear saturation density.  I also give some discussions about
the substantiation of quark matter by means of the gravitational wave
signals including the multi-messenger prospect.
\end{abstract}

\maketitle

\section{Important lessons we have learned from the heavy-ion 
  collision}

It is well-known that quantum chromodynamics (QCD) has two extreme
limits that exhibit global symmetries to characterize the QCD phases.
These symmetries are only approximate in the realistic case with physical parameters.
In the limit of zero quark mass, chiral symmetry becomes exact, which
is spontaneously broken at low energy as
\begin{equation}
  SU(\Nf)_{\mathrm{L}} \times SU(\Nf)_{\mathrm{R}} ~\to~
  SU(\Nf)_{\mathrm{V}}\,,
  \label{eq:sym_chiral}
\end{equation}
where $\Nf$ is the number of quark flavors.  In practice, $\Nf$ varies
according to the typical energy scale of relevant processes as
compared to the quark masses.  For the environment in the neutron
star, the quark chemical potential is less than $1\GeV$, so the
active quark flavors are limited to the up, the down, and the strange.
This corresponds to an intermediate situation between $\Nf=2$ and
$\Nf=3$, which is often referred to as $\Nf=2+1$.  In the opposite
limit of infinite quark mass or the quenched limit of $\Nf=0$, center
symmetry becomes exact, which is realized in the low-temperature phase
and spontaneously broken in the high-temperature phase as
\begin{equation}
  Z_{\Nc} ~\to~ \mathbb{1}
  \label{eq:sym_deconf}
\end{equation}
with the color number $\Nc$ that is $\Nc=3$ in QCD but could be
artificially adjusted for theoretical purposes.  If the realization of
the global symmetries differs as physical parameters change, there
must be at least one phase transition at some critical values of the
parameters; For the temperature as the physical parameter, the
critical value is denoted by $\Tc$.  As a function of
the quark mass $\mq$, two phase transitions must be found at $\mq=0$
and $\mq=\infty$.  Unfortunately, however, the above global symmetries
cannot coexist for any $\mq\neq 0$.  The question is which phase
transition is closer to the physical point.

This question has been intensely investigated in the context of the
search for the quark-gluon plasma (QGP) at high temperature and low
baryon density.  Here, I only summarize what we know about the natures
of the QCD phase transitions without going into phenomenological
details about how they have been concluded.
\vspace{1em}

\noindent  
\emph{--- The chiral phase transition has not been observed.}

The masses of the up and the down quarks are two orders of magnitude
smaller than the typical QCD scale, i.e., $\LQCD\simeq 0.2\GeV$.
Therefore, it would be conceivable that the physical point may be not
so far from the chiral limit of $\mq\to 0$, and if so, some remnant
phenomena associated with chiral symmetry
breaking~\eqref{eq:sym_chiral} should be expected.  Strangely,
however, no criticality has been confirmed by the experimental data
from the heavy-ion collision.  For this missing link between chiral
symmetry and the heavy-ion collision data, there are three (and
probably more) explanations.

The first one is that the critical region is too small.  Although
$\mq\ll \LQCD$ is certainly the case, the physical pion mass is
comparable with the QCD scale, i.e., $m_\pi \simeq \LQCD$, and the
soft mode cannot be sufficiently soft around the pseudo-critical
temperature.  It should be noted that the pseudo-scalar channel is the
lightest if the density is zero and the QCD inequality is
applied~\citep{Nussinov:1999sx}.  The second explanation is that
chiral symmetry affects hadron spectra, but does not drastically
change the physical degrees of freedom, which implies that the bulk
thermodynamics is not sensitive to the chiral phase transition.  To
observe consequences from the chiral phase transition, thus, some
excitation on top of the bulk thermodynamics such as fluctuations and
transport properties must be measured, but such quantities are rather
fragile and easily diluted in the average over the whole dynamical
evolution.  The third problem is that good probes for chiral symmetry
are not quite established, in my opinion.  I know that people have
discussed the medium modification of the vector mesons quantified by
the dilepton measurements.  Theoretically speaking, it is correct to
say that, once the vector meson spectral function is somehow obtained,
the expectation values of operators are constrained by the QCD sum
rule.  Even if the vector meson mass decreases, however, it does not
uniquely lead to the decreasing behavior of the chiral order
parameter; in other words, the Brown-Rho scaling is a convenient
hypothesis but not a theorem.  For a counter-example,
see~\citep{Klingl:1997kf}.  The vector meson may exhibit not the mass
shift but only the width broadening, and then the connection to chiral
symmetry is more subtle.
\vspace{1em}

\noindent 
\emph{--- The deconfinement phase transition has not been observed.}

Then, one may want to think that the transition associated with the
symmetry change~\eqref{eq:sym_deconf} should have been probed, but it
is not the case, either.  In the absence of dynamical quarks, the pure
gluonic theory has a first-order phase transition for $\Nc=3$, but
dynamical quarks smear center symmetry of $Z_{\Nc}$.  As a result,
there is no sharp phase transition of deconfinement at all, while the
thermodynamic quantities such as the energy density grow up within a
narrow range of the temperature.  In the past, some people were
serious about the phenomenological implication from the symmetry
breaking~\eqref{eq:sym_deconf}, that is, the dynamics of the QCD
center domain walls associated with the discrete symmetry breaking in
the early universe and the primordial magnetic field.  Nowadays,
however, it is a well-established fact that the deconfinement is a
smooth phenomenon and there is no hope to salvage any hint about the
center domain walls associated with deconfinement.

The absence of any sharp phase transition of deconfinement also
implies a radical conjecture.  Suppose that QCD has strict color
confinement at zero temperature, then, what should be the probability to
find quark excitation at low temperature, say, $\sim 0.01\GeV$?   Is
it strictly zero as literally imposed by color confinement or is it
exponentially suppressed and yet nonzero?  When I was a young postdoc,
I asked this question to a renowned physicist and the answer I got was
quite disappointing.  I can never forget his answer --- I don't trust
the finite-temperature formalism, that is why I wrote papers on the
system in finite volume box, but not a single paper at finite
temperature...  At that time, I was young enough to stave off
pessimism, and I kept my face up to write finite-temperature papers.
Anyway, in spite of long-standing efforts, the order parameter for
deconfinement has not been discovered except for the pure gluonic
case.  More precisely speaking, only the quark deconfinement order
parameter (the Polyakov loop in the fundamental representation) is
established and even in the pure gluonic theory the gluon
deconfinement order parameter is not known or it does not exist;  the
Polyakov loop in the adjoint representation always takes a nonzero
value.  It would be not a bizarre idea to interpret the absence of the
order parameter as the absence of phase transitional changes at all.
Then, as soon as a finite temperature is introduced, the thermal
excitation of colored objects may not be prohibited stringently.  If
this statement makes sense, the definition of the QGP would be
obscure.
\vspace{1em}

\noindent
\emph{-- The crossover is boring, but it is the real, and the most
  difficult to understand:~}

Theoreticians tend to idealize the system taking the limit of either
$\mq=0$ or $\mq=\infty$.  Critical phenomena are characterized by
model independent analyses in theory and the experimental challenge
to probe them is very appealing.  However, the physical quark mass
is not close to these limiting values, and the change from hadronic
matter to the QGP takes place only smoothly if the baryon density is
small.  This smooth change is commonly referred to as the
``crossover'' in the QCD community.  This usage of the terminology is
analogous to that of the BEC-BCS crossover.

The scenario of no phase transition may sound like a boring reality,
and indeed, this is the most boring in the phenomenological sense.
There is no critical enhancement of fluctuations at the second-order phase
transition, and there is no bubble nucleation at the first-order phase
transition.  There is no clear signature for the realization of the
QGP, unfortunately.

\begin{figure}
  \centering 
  \includegraphics[width=0.6\columnwidth]{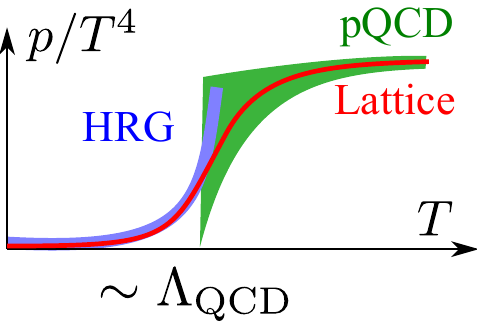}
  \caption{Schematic picture of the crossover from hadronic matter to 
    the QGP.  The figure is adapted from \citep{Fukushima:2014pha}.}
  \label{fig:trans}
\end{figure}

Even though there is no exact way to define the QGP in quantum field 
theory, one could reasonably declare the creation of the QGP if any 
hadronic calculations cannot reproduce observed properties and 
perturbative QCD (pQCD) results can be consistent with them.  The idea
is visualized in Fig.~\ref{fig:trans}.  Actually, I emphasized this
picture of the crossover in the proceedings contribution to
\emph{Quark Matter 2014}~\citep{Fukushima:2014pha}.  In the
low-temperature phase, the hadron resonance gas (HRG) model agrees
with the thermodynamic quantities measured in the lattice-QCD
simulation, but the pressure from the HRG model eventually blows up.
From the high-temperature side, the resummed perturbation theory of
QCD works very well to explain the lattice-QCD thermodynamics up to
some temperature, but the uncertainty is getting larger at lower
temperatures.  In this way, one can distinguish the hadronic phase
well approximated by the HRG model and the QGP well described by the
pQCD-based calculation.

One may wonder how these distinct regimes could be smoothly connected.
In fact, that is the question.  There are several preceding attempts
along these lines.  For example, the validity region of the HRG model
can be extended toward higher temperature by the inclusion of the
excluded volume effects~\citep{Andronic:2012ut}.  Also, the resummed
pQCD results near the pseudo-critical temperature may be ameliorated
by the inclusion of the random gauge background modeled by the
Polyakov loop averaging~\citep{Vuorinen:2006nz}.  For some extension in
terms of the semi-QGP, see~\citep{Hidaka:2008dr}.  Then, it is likely
that the improved hadronic and pQCD results may have the overlapping
region of validity, in which one may say that matter contents are both
hadronic and partonic (quarks/gluons).  I must judiciously emphasize
that I am not talking about a mixture of 
hadrons and partons.  The identical physical system near the
pseudo-critical temperature is equivalently regarded as either
hadronic matter or partonic matter, and this concept is nothing but
the duality in physics.

Although it is boring in phenomenology, the crossover as realized by
the duality is the most difficult to understand from the microscopic
mechanism.  This makes a contrast to a theoretical description of
quark hadron mixture; see, e.g., \cite{Blaschke:2023pqd} in a
Beth-Uhlenbeck approach.  In contrast, in the duality region of the
temperature, two descriptions in terms of hadrons and partons
marginally work, and in turn, they nearly break down.  Thus, besides
the numerical simulations, there is no reliable modeling of the
duality to investigate the matter properties in this intermediate
region (however, see \cite{Fujimoto:2023mzy} for a recent
development).  Nevertheless, if one is interested in the
thermodynamics only without caring about microscopic ingredients, one
can construct it by a smooth interpolation between the HRG and the
pQCD branches, as depicted in Fig.~\ref{fig:trans}.
See also \cite{Albright:2014gva} for a hybrid matching prescription.

\section{Not the best but the better among worse at high baryon 
  density}

We live on the first-order phase transition.  Nuclei are self-bound
fermionic systems.  If the baryon chemical potential, $\muB$, is too
small, no baryonic matter can exist at zero temperature.  With
increasing $\muB$ in isospin symmetric matter, a first-order
phase transition occurs and the baryon density jumps from zero to the
saturation density.  This is a phase transition from the vacuum to a
liquid of nuclear matter.  Of course, we cannot live in a liquid of
nuclear matter at the saturation density.  So, our world lies exactly
at the onset, $\muB=m_N-B$, where $m_N$ is the lightest baryon
(nucleon) mass and $B$ is the binding energy $\sim 0.016\GeV$ per
nucleon.

The first-order phase transition is expected whenever such self-bound
fermionic systems emerge.  Therefore, if quark droplets are stable (or
meta-stable), which hints the existence of the quark star, a
first-order phase transition may be found at higher densities than the
nuclear saturation density.  This fascinating possibility cannot be
ruled out so far, and the quark star hunt is one of the most exciting
activities in nuclear and astro physics.

\begin{figure}
  \centering 
  \includegraphics[width=0.98\columnwidth]{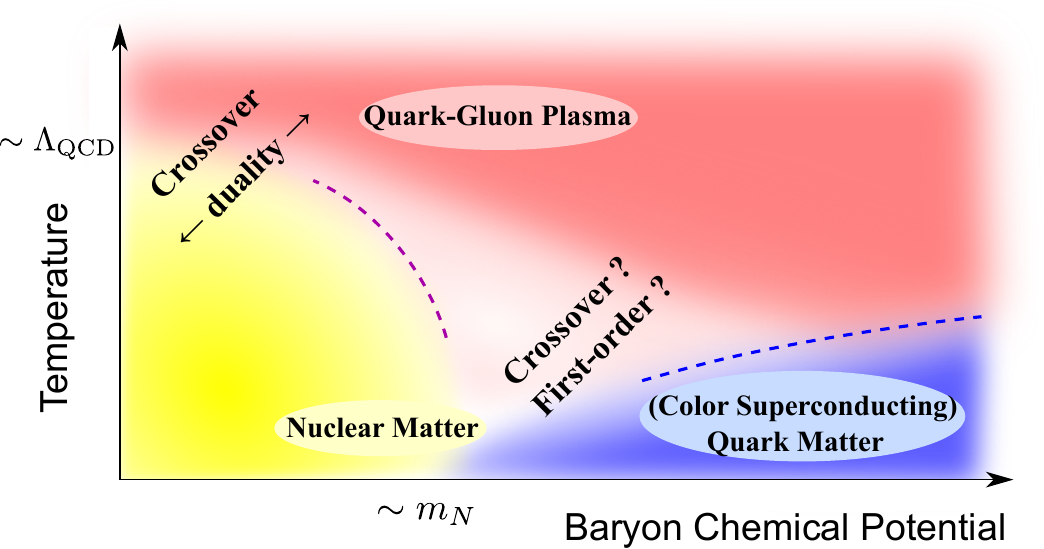}
  \caption{Crossover vs. first-order phase transition scenarios at 
    high baryon density on the QCD phase diagram.  The figure is 
    adapted from \cite{Fujimoto:2022xhv}.}
  \label{fig:phase}
\end{figure}

Now, let us recall the lessons from the heavy-ion collision.  In the
past, many intriguing possibilities were considered, but after all,
the most boring scenario turned out to be the case.  Based on the
wisdom we earned from high-temperature QCD matter, I would say, it
should not be too conservative even if I postulate the least
structured phase diagram, namely, only the crossover from hadronic
matter to quark matter at high densities; see Fig.~\ref{fig:phase} for
scenarios of the crossover and the first-order phase transition on the
phase diagram.  Because the duality plays such an essential role for
deconfinement at high temperature, we must ask to ourselves; why not
at high baryon density?  Such an assumption of the least structured
phase diagram may sound like the well-known principle of Occam's
razor, but I must say, the duality of deconfinement is a concrete
physical concept and the crossover should conceal rich physics behind
it.
For possible theoretical justification, let me refer to the
quark-hadron continuity scenario between a superfluid strange nuclear
matter and a three-flavor color-superconducting
matter \citep{Schafer:1998ef,Alford:1999pa,Hatsuda:2006ps}, which is
extendable to the two-flavor case \citep{Fujimoto:2019sxg}.  However,
recent discussions of placing a superfluid vortex as a probe have
revealed non-trivial topological nature of quark
matter \citep{Chatterjee:2018nxe,Alford:2018mqj,Cherman:2018jir}.

It is still an attractive argument that quark matter might
energetically be favored after a first-order phase transition, which
is a stereotype scenario based on the bag model that provides us with an
intuitive insight into the deconfining mechanism.  However, again, we
must ask to ourselves; if the bag model picture were validated at high
density, why not at high temperature?  It is already the fact that the
QCD phase transition at high temperature and low baryon density is a
smooth crossover and the duality connects two regimes in terms of
different physical degrees of freedom.  From this modern view of
deconfinement, I believe that the crossover scenario at high density
should be rather natural than others that require too specific assumptions,
whereas the crossover scenario may not hit the exact ground truth.  I
remember that when I was a student at junior high school, an art
teacher told us how to make a dessin, that is a rough sketch of
drawing, of a human body for example.  Then, first, one should be
careful of the whole body and the balance of volumes and shapes of the
body parts.  It is not recommendable to draw full details of one arm
only before placing the shoulder, the head, etc.  It is sometimes
perplexing to me that some people stick to minutiae such as an
undenied possibility of weak first-order phase transitions before
taking a whole perspective of the QCD phase diagram.  Probably, a fine dessin
of hands and fingers inveigles some theoreticians.

\section{Likely EoSs supporting the crossover but the first-order 
  phase transitions popping up}

I am not so negative against the scenario of the first-order phase
transition.  I am simply saying that the crossover scenario is a good
starting point to draw the first approximation for the equation of
state (EoS), and then one can engrave this base with more structures.
To make my point clear, let us take a close look at the inferred EoSs
from the machine-learning analysis.  We have invested resources to
establish a method to solve the inverse problem from the neutron star
data on the radius-mass ($R$-$M$) plane to the EoS candidate.  The
technique was first developed by means of the Bayesian analysis
\citep{Steiner:2010fz,Ozel:2015fia,Annala:2019puf,Brandes:2022nxa}.
It is extremely nontrivial in this method of the Bayesian analysis how
to evaluate the conditioned probability to find the observation data
for the given EoS, that is, the likelihood.  Once the EoS is
specified, one line of the combination of $(R,M)$ of the neutron star
is fixed.  Then, suppose that this $(R,M)$ line is the true answer,
how can we quantify the probability to obtain the observation data
with error bars? Actually, the distribution of the observation data
may be significantly different from the Gaussian, and there is no
simple machinery to quantify the conditioned probability.  Therefore,
some sort of the Monte-Carlo simulation is indispensable, and the
entire procedures are time consuming.

\begin{figure}
  \centering 
  \includegraphics[width=0.9\columnwidth]{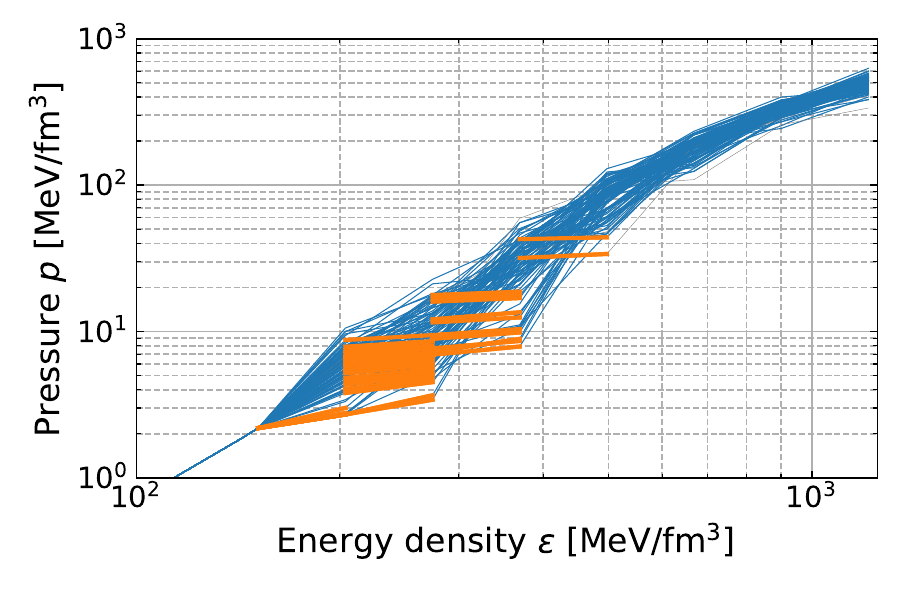}
  \caption{100 EoSs inferred from 100 independently trained models.
    The input is the common real observation data of the neutron star
    radius and mass.  The orange-colored plateau represents the
    approximately identified region of the first-order phase
    transition.  The figure is adapted from \citep{Fujimoto:2021zas}.}
  \label{fig:first}
\end{figure}

Another machine-learning technique that utilizes the neural network
demonstrates a good performance to predict an EoS candidate in
response to a given observation dataset.  AI allergic people say that
the machine-learning method is a black box and the Bayesian analysis
is superior.  But we should keep in mind that the Bayesian analysis is
one of the useful machine-learning methods, and we can choose the most
convenient method for the problem we want to solve.  Our final
objective is to quantify the posterior distribution of the EoS, and we
can achieve this goal using the neural network approach as follows.
In the supervised learning, the model is trained by the training data,
and the trained model predicts the output EoS for the input data.  We
can prepare independent trained models with independent training
data.  If the output is not well constrained, different models should
lead to different predictions for the same input, and vice versa.  We have
conducted this numerical experiment~\citep{Fujimoto:2021zas} and the
results are shown in Fig.~\ref{fig:first}.

Among 100 outputs from independently trained models, in
Fig.~\ref{fig:first} we see that some flat regions occasionally
appear.  The orange lines indicate the regions with the slope, $c_s^2
= dp/d\varepsilon$, less than 0.01, which can approximately be
regarded as the first-order phase transition.  If the EoS is
parametrized by the piecewise polytrope with more segments, weaker
first-order phase transitions can be more easily accommodated.  The
general tendency is that the lower density region may have a better
chance to have the first-order phase transition, but the
interpretation needs caution.  For the piecewise polytrope, it is
customary to make the partitions with equal interval in the
logarithmic scale, and this means that the first-order phase
transition in Fig.~\ref{fig:first} gets stronger at higher density by
construction of the piecewise polytrope.  Therefore, naturally, the
stronger first-order phase transitions at higher densities are less
favored.

Here, a message I want to convey from Fig.~\ref{fig:first} is not the
relevance of the weak first-order phase transition but my view of the
EoS construction;  the overall behavior of the EoS is the averaged one
which is very smooth and there is no characteristic point in the
averaged global shape.  Then, within the uncertainty band estimated by
fluctuating outputs in Fig.~\ref{fig:first}, the EoS can gradually be
updated by future data.  Hereafter, I discuss the averaged EoS only
but I never exclude the possibility of weak first-order phase
transitions.  Simply, the currently available data are not sufficient
to resolve such ripples on top of the bulk EoS base.

\begin{figure}
  \centering 
  \includegraphics[width=0.9\columnwidth]{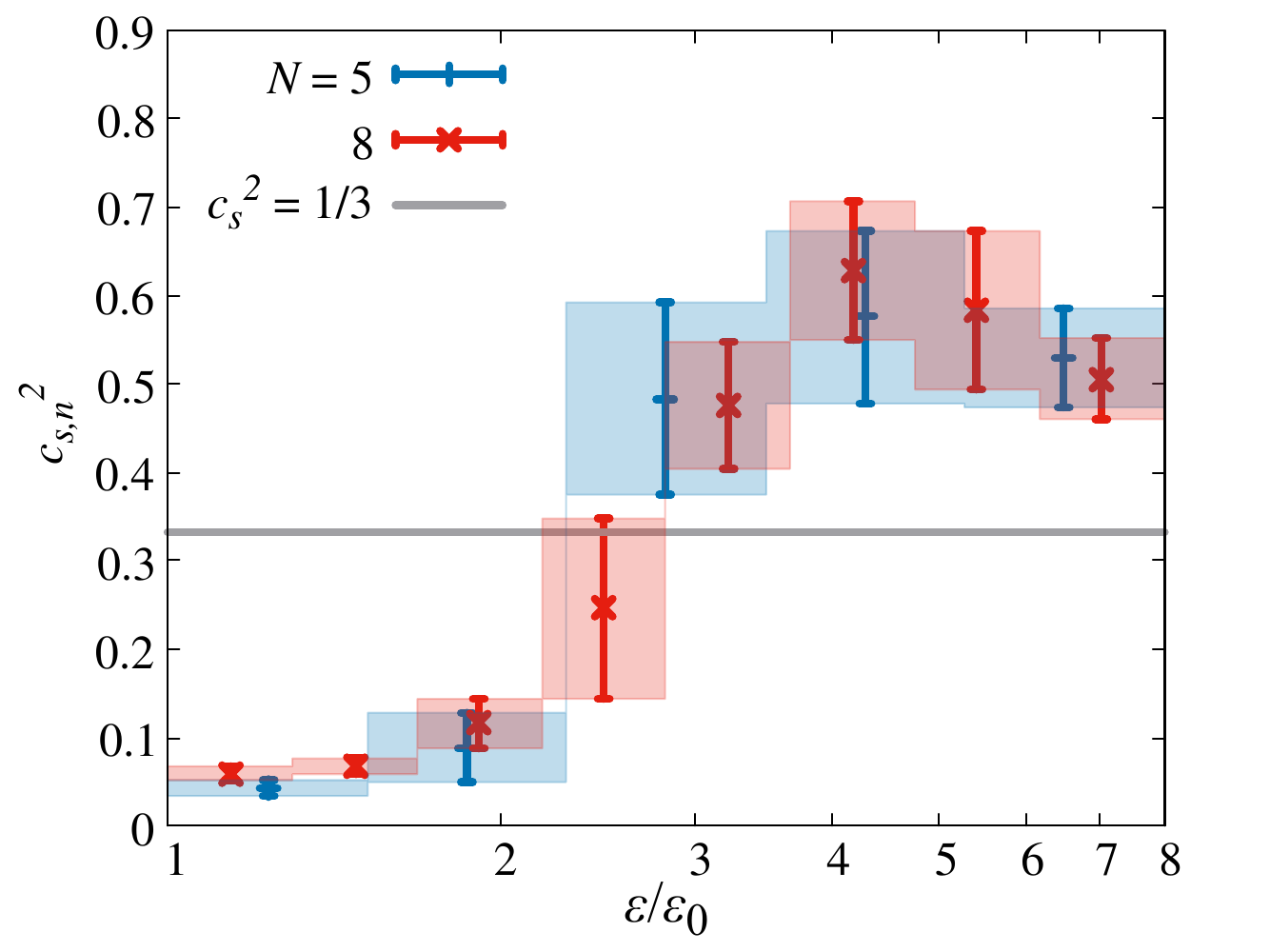}
  \caption{Speed of sound squared as a function of the energy density 
    in the unit of the energy at the saturation density, namely, 
    $\varepsilon_0=0.15\GeV/\mathrm{fm}^3$.  The dependence on the 
    number of segments, $N$, is only minor.  The figure is adapted 
    from \cite{Fujimoto:2024cyv}.}
  \label{fig:sound}
\end{figure}

The slope of the EoS is the speed of sound squared, i.e.,
$c_s^2=dp/d\varepsilon$, and this quantity approaches $1/3$ in the
conformal limit.  If there is no mass scale other than the chemical
potential, the pressure is parameterized with a numerical coefficient as
\begin{equation}
  p = \nu \muB^4\,,
  \label{eq:p4}
\end{equation}
and then the energy density is
$\varepsilon = \muB(dp/d\muB) - p = 3\nu\muB^4$.  Thus, one can
immediately conclude $c_s^2=1/3$ in this case.  The latest results of
the speed of sound from the machine-learning inference are presented
in Fig.~\ref{fig:sound}.  Although the error bar (the $1\sigma$ band)
is not small, it is fair to say that a peak may stand around
$\varepsilon \simeq 4\varepsilon_0 \simeq 0.6\GeV/\mathrm{fm}^3$.  I
am a bit prudent about the statement.  I was once sloppy enough to
give a talk and use a word, peak, by showing Fig.~\ref{fig:sound}.
Then, an experimental colleague in the audience immediately reacted to
refute my careless statement and said that no peak was visible.  It was
experimentally a correct remark, but as a theoretician, I would still
insist that this should be a peak even though the peak width might be
broad.  Since QCD is an asymptotically free theory, the speed of sound
squared must go to the conformal value, $1/3$, eventually in the high
density limit.

It is actually a nontrivial question whether the slightly decreasing
behavior for $\varepsilon/\varepsilon_0 > 4$ is physical or
artificial.  In the older analysis in \citep{Fujimoto:2021zas}, we were
inclined to consider that the decrease was attributed to the
machine-learning artifact.  If there is no constraint from the data,
the output should be inherited from the distribution of the speed of
sound squared in the training data, that is,
$c_s^2\sim 0.5$.  However, in the follow-up analysis in
\citep{Fujimoto:2024cyv}, we have confirmed that the decreasing
behavior even for $\varepsilon/\varepsilon_0 > 4$ is well constrained
by the observation
data.  This sounds neat; however, this is a very puzzling assertion.
From the astrophysical constraint, the soft EoS
can be excluded by the presence of massive neutron stars, but the
stiffer EoSs are always allowed as long as it does not violate the
causality bound.  Physically speaking, there is no reason why $c_s^2$
should decrease for $\varepsilon/\varepsilon_0 > 4$.  Among the
authors of \citep{Fujimoto:2024cyv}, we had a lot of discussions, and
our hypothetical conclusion is that the machine-learning model might
give an additional meaning to the ``absence'' of further massive
neutron stars.  From the discovery of something, usually, people can
impose a constraint so that this something can exist.  But, from the
non-discovery, we have no idea whether something does exist but it is
not discovered yet, or this something does not exist at all.
Therefore, we usually do not care about the non-discovery, but the
machine-learning model does.  Actually, it is not an insane idea to
take the non-discovery such actively.  In science, some principles,
desirably simpler principles, are assumed and they are accepted unless
some counter-example is found.  In the same way, we may well assume
that the non-discovery signifies the non-existence.  Under this
assumption, the heaviest neutron star is to be identified as the upper limit
and the EoS cannot be stiff to exceed this upper limit.  One may say
that this is just an artifact from the machine-learning inference, but
I myself was thrilled about the potential of the machine-learning
approach that may break out of our shell of thinking.

\section{Trace anomaly approaching zero means what?}

Once the EoS is somehow inferred, various combinations of $p$ and
$\varepsilon$ can be considered, and the speed of sound squared is
only one example.  Another interesting quantity is the conformality
measure defined by
\begin{equation}
  \Delta = \frac{1}{3} - \frac{p}{\varepsilon}
  = \frac{\varepsilon - 3p}{3\varepsilon}\,.
  \label{eq:conformality}
\end{equation}
The numerator, $\varepsilon - 3p$, is the trace of the energy-momentum
tensor.  The energy-momentum tensor and the dilatation current can be
improved so that the divergence of the dilatation current can be
expressed by the trace of the energy-momentum tensor.  Therefore, if
the dilatation current is conserved due to conformal symmetry, the
trace of the energy-momentum tensor is vanishing.  In the classical
level QCD is an almost scale free theory except for the quark masses,
but the running coupling involves an additional scale, $\LQCD$, from
the quantum effects.  Then, the trace of the energy-momentum tensor is
no longer vanishing but it is given by a combination of the gluon
condensate and the $\beta$ function that characterizes the running
coupling.

Here again, let me explain what is known about the trace anomaly at
high temperature.
\vspace{1em}

\noindent
\emph{--- The trace anomaly in the matter part is peaked around the 
  critical temperature.}

In lattice QCD, the dimensionless quantity, $(\varepsilon - 3p)/T^4$,
is often called the interaction measure.  In the pure gluonic
theory~\citep{Boyd:1996bx} as well as QCD, a peak in the interaction
measure stands near the critical temperature.  It would be misleading
to say that the trace anomaly is enhanced near the critical point,
though.  The fact is rather opposite.  The vacuum part of the gluon
condensate is positive, and the gluon condensate melts around the
critical temperature, which means that the negative matter part
partially cancels the positive vacuum part.  Then, for the
temperatures higher than $\Tc$, the gluon condensate goes to
negative (the trace anomaly is positive).  Therefore, the correct
interpretation is that the trace anomaly including the vacuum part is
not enhanced but crossing the zero near the critical temperature.

At higher temperatures, one may think that the interaction measure
should be suppressed as $T^{-4}$, but this is not really the case.  It
was recognized that the interaction measure scales as $T^{-2}$ at high
temperature, which implies that the finite-temperature gluon
condensate has a term behaving nonperturbatively like $\LQCD^2 T^2$
even at high temperature.

\begin{figure}
  \centering 
  \includegraphics[width=0.9\columnwidth]{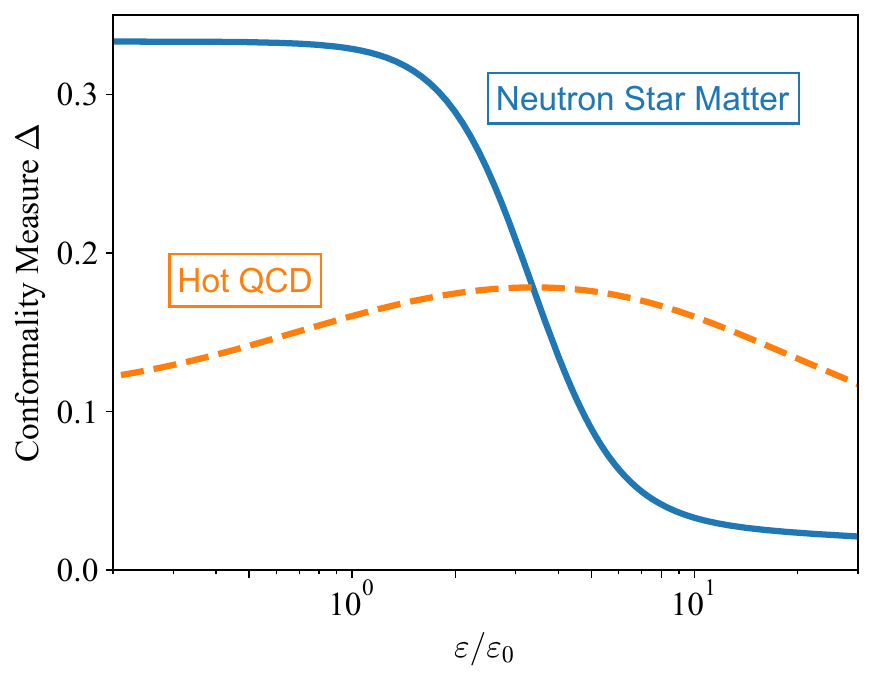}
  \caption{Conformality measure from the neutron star data (solid 
    line) and the finite-temperature lattice-QCD data (dashed line). 
    The horizontal axis is the energy density in the unit of 
    $\varepsilon_0=0.15\GeV/\mathrm{fm}^3$.}
  \label{fig:conformal}
\end{figure}

The pressure, $p$, as a function of the temperature, $T$, is parametrized in
\citep{HotQCD:2014kol} based on the first-principles calculation, and
it is easy to compute the conformality
measure~\eqref{eq:conformality}.  In the QCD case with dynamical
fermions, the peak is smeared as shown by the dashed line in
Fig.~\ref{fig:conformal}, while a sharp peak appears in the pure
gluonic theory.

A surprise awaits to be found when we consider $\Delta$ using the
neutron star data.
\vspace{1em}

\noindent 
\emph{--- The trace anomaly in the matter part rapidly decreases in 
  the central cores of the neutron star.}

The solid line in Fig.~\ref{fig:conformal} shows $\Delta$
corresponding to the neutron star EoS in Fig.~\ref{fig:sound}, which
is parametrized in \citep{Fujimoto:2022ohj}.  I would point out that
there are (at least) two surprises.  One is that the characteristic
energy scales are surprisingly small and they are relevant to the
realistic neutron star environment.  We see that $\Delta$ rapidly
decreases already around $\varepsilon/\varepsilon_0 \gtrsim 2$.  Then,
around $\varepsilon/\varepsilon_0 \sim 5$ which is reachable in the
deepest core of the heavy neutron star, the conformality measure is
strongly suppressed and, it could even go to negative, which is the
second surprise.

At finite density and low temperature, the conformality measure starts
from $\sim 1/3$ in the low-density side, and this can be understood
from $\varepsilon \gg p$ in the nonrelativistic regime.  In other
words, along the finite density direction, the nucleon mass
significantly breaks scale invariance and $\Delta$ is naturally
enhanced there.  Unlike the finite temperature case, it is subtle to
interpret the behavior of this matter part in terms of the gluon
condensate.  One may say that the trace anomaly has not only
the gluon condensate but also the chiral condensate, but the latter
disappears in the massless quark limit.  I believe that the
qualitative and even the quantitative nature of $\Delta$ as shown in
Fig.~\ref{fig:conformal} would hardly be changed in the massless
limit.  Na\"{i}vely in the QCD language, together with the vacuum
part, the positive gluon condensate seems to increase further as the
density grows up.  This is unnatural at first glance, but this
seemingly unnatural behavior could be justified if some forms of new
condensates develop with increasing density.

The necessity of additional condensates is strongly motivated by the
following argument as well.  If $\Delta$ goes to negative, it is
straightforward to see,
\begin{equation}
  \varepsilon - 3p = \muB^5 \frac{d}{d\muB} \frac{p}{\muB^4} < 0\,,
\end{equation}
which means $d\nu/d\muB < 0$ where $\nu$ is the thermal degrees of
freedom introduced in Eq.~\eqref{eq:p4}.  This is an incredible
statement;  the thermal degrees of freedom decreases with increasing
chemical potential.  How can it be true?  Interestingly enough,
there are several known examples in which $\varepsilon - 3p < 0$ is
realized.  The most well-known system is the high isospin matter with
pion condensation~\citep{Son:2000xc}.  The chiral perturbation theory
should work as long
as the isopin chemical potential is not too large, and
$\varepsilon - 3p< 0$ is predicted, which has been confirmed by the
lattice simulation~\citep{Abbott:2023coj}.  It is interesting to point out that the
nonperturbative quadratic term such as $f_\pi^2 \mu_{\mathrm{I}}^2
\sim \LQCD^2 \mu_{\mathrm{I}}^2$ is
indispensable to account for the enhanced behavior of the speed of
sound~\citep{Chiba:2023ftg} and also for the negative conformality
measure.  At high temperature
and high density, respectively, the manifestation of $\Delta$ looks
very different, but in both cases the nonperturbative quadratic term
$\sim \LQCD^2 T^2$ or $\sim \LQCD \mu^2$ should be required.  The
microscopic origin might be common.

In the neutron star matter, it is likely that $\Delta$ crosses zero
and $\Delta < 0$ is realized around
$\varepsilon/\varepsilon_0 \simeq 4$-$5$.  The promising candidate for
the condensate there is the color-superconducting gap.  Because the
coupling constant is too small, the effect of the color superconductor is only
moderate at high density where the pQCD treatment is
justified~\citep{Fukushima:2024gmp}.  For the consistent
interpretation of the neutron star EoS, nonperturbative
enhancement of some condensates must be taken into account, which is a
challenging problem.

\section{Scanning the phase transitions with the total mass of the binary neutron star
  system}

The merger of the binary neutron stars is such an explosive event that
the future gravitational wave measurement is expected to give us
useful information about extremely dense matter even for
$\varepsilon \gtrsim 5\varepsilon_0$.  Also in the intermediate
density region, $\varepsilon \lesssim 5\varepsilon_0$, the tidal
deformability in the inspiral stage before the merger can constrain
the EoS well, and the uncertainty band in Fig.~\ref{fig:first} should
become much narrower soon.
This is actually why I tend to claim that the fine details of the EoS
should be updated later when more experimental data will be available,
which will happen in the near future.

\begin{figure}
  \centering 
  \includegraphics[width=0.9\columnwidth]{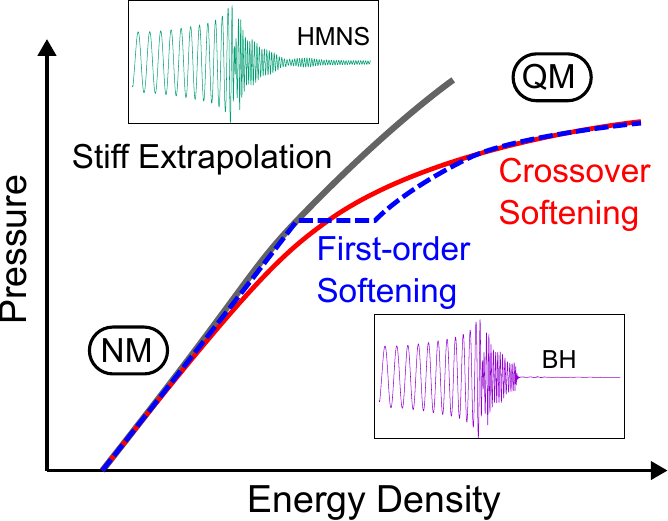}
  \caption{Schematic illustration of the phase transition scan by the
    gravitational wave signals.  The softening from nuclear matter
    (NM) to quark matter (QM) changes the life time of the remnant
    after the merger.  If the EoS is stiff, the hypermassive (or
    supramassive) neutron star (HMNS) remains, while the soft QM EoS
    cannot support it and the gravitational collapse to the blackhole
    (BH) is expected.}
  \label{fig:schematic}
\end{figure}

It is the most ambitious attempt to utilize the gravitational wave
signals for the purpose of the detection of the QCD phase transitions.
For the moment, nobody knows the exact phase structure at high
density, and my belief
is the crossover from hadronic to quark matter as emphasized
previously.  As a matter of fact,
the order of the phase transition itself is not so important in this
context.  If the pQCD estimate of the EoS is extrapolated toward the
neutron star
density, the pQCD-based EoS is very soft, and the speed of sound
squared cannot be far from the conformal value.  For the fate of the
binary neutron star merger, it is essential that the EoS of dense
matter after some phase transition or crossover is expected to be
soft, and it is almost irrelevant how to transmute the matter with or
without a discontinuous jump.  The bottom line is the
following:  the QCD phase transition, which may be a first-order phase
transition~\citep{Most:2018eaw, Bauswein:2018bma, Weih:2019xvw} or a
continuous crossover~\citep{Fujimoto:2022xhv}, softens the EoS and the
gravitational collapse to the blackhole is triggered by the onset of
quark matter.  The idea is schematically illustrated in
Fig.~\ref{fig:schematic}.  If the full shapes of the gravitational
waves from the post-merger stage will be probed in the future, they
would provide us with valuable information about the nature of the
phase transitional change to quark matter (for example, the shift of
the peak frequency as discussed in \cite{Bauswein:2018bma}).  However,
the nature is not so kind.  After the merger, the system size is far
smaller than the inspiral stage, and the amplitude of the
gravitational waves is much reduced.  The typical frequency range of
the gravitational waves of our interest to investigate the QCD phase
transition is around
$\gtrsim 3\,\mathrm{kHz}$ and I am not very optimistic about the
possibility to achieve the sensitivity of the detector in this
frequency range.  To confirm the QCD phase transition using the
post-merger gravitational wave signals, we should be very lucky or we
should complete the mission impossible to make a reliable and precise
prediction or we should wait for decades till the detector sensitivity
after several generations will be sufficiently improved.

\begin{figure}
  \centering 
  \includegraphics[width=0.9\columnwidth]{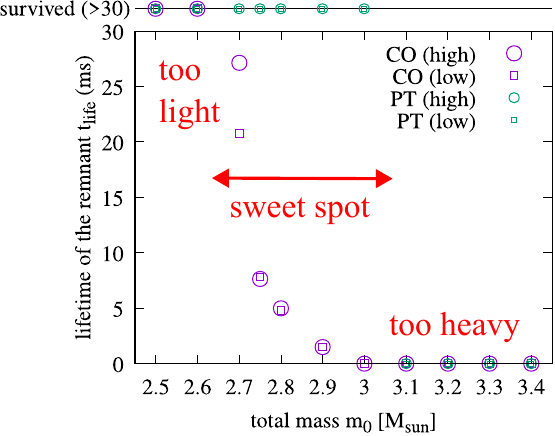}
  \caption{Life time of the remnant after the merger as a function of 
    the total mass of the binary neutron star system.  CO represents 
    the intermediate-density crossover scenario and PT represents the 
    high-density first-order phase transition scenario.  The figure is 
    adapted from \citep{Fujimoto:2024ymt}.}
  \label{fig:life}
\end{figure}

One workable breakthrough to overcome the sensitivity problem at high
frequency band is the
multi-messenger analysis.  In the case of GW170817, an electromagnetic
counterpart, AT2017gfo, identified as the kilonova has been observed.
The extensive analysis of the kilonova gave us a lot of information
about the r-process elements~\citep{Domoto:2021xfq}.  Here, I want to
emphasize that the presence of the kilonova already imposes a
condition onto the EoS candidate.  The point is that the life time of
the remnant after the merger should be long enough to acquire the disk
mass in a way consistent with the estimated kilonova mass.  Any
phase transition scenario that cannot explain the kilonova
mass must be strictly ruled out.  If the luminosity of many observed
kilonovae will be systematically investigated in the future, the phase
transition scenarios can be discriminated even without looking at the
post-merger signals.  In fact, Fig.~\ref{fig:life} shows the life time of the
remnant after the merger as a function of the total mass of the binary
neutron star system~\citep{Fujimoto:2024ymt}; see also the related
discussions on the threshold mass in \citep{Bauswein:2020aag}.  CO and
PT are two representative EoS scenarios.
It is evident that the life time strongly depends on the EoS
scenarios in a specific window of the total mass (labeled as ``sweet
spot'' in Fig.~\ref{fig:life}).  For example, according to
Fig.~\ref{fig:life}, if a bright kilonova will be discovered for the
merger with the total mass at $2.9M_{\odot}$, then the CO scenario
must be denied.  We note that, if the binary system is too light, the
hypermassive neutron star lives long enough and the EoS differences
are irrelevant.  Also, if the binary system is too heavy, the remnant
is immediately collapsed into the blackhole regardless of the EoS
differences.  The sweet spot is centered around twice of the typical
neutron star mass $\sim 2.8M_\odot$, which was indeed the case for
GW170817.  In this way, the location and the strength of the
phase transition including the crossover will be able to be
constrained by the future observation in the total mass window of the
sweet spot.  In some events, the kilonova might be simply overlooked
experimentally, but not always so.  I am very optimistic about the
systematic survey of the kilonova which will be able to render a
verdict of the quark matter transition.

\section*{Acknowledgments}
This proceedings contribution is based on my published papers as well
as ongoing projects.  The author thanks his collaborators of the
published papers:
Yuki~Fujimoto, Kenta~Hotokezaka, Syo~Kamata, Koutarou~Kyutoku,\\
Larry~McLerran, Shuhei~Minato,
Koichi~Murase, and\\
Michal~Praszalowicz.
He also thanks his collaborators of the ongoing projects:
Len~Brandes, Kei~Iida, and Chengpeng~Yu.

\bibliographystyle{cas-model2-names}
\bibliography{fukushima}

\end{document}